\documentclass[aps,prd,twocolumn,floatfix,nofootinbib,superscriptaddress]{revtex4-1}
\usepackage{dcolumn}

\usepackage{amsmath,amsfonts,extarrows,amssymb,mathrsfs,times,bm}
\usepackage{extarrows}
\usepackage{graphicx}
\usepackage{float}
\usepackage{graphicx,epstopdf}
\usepackage{dcolumn}
\usepackage{exscale}
\usepackage{relsize}
\usepackage{mathtools}
\usepackage{xcolor}
\usepackage{mhchem}
\usepackage[colorlinks=true,breaklinks=true,linkcolor=blue,citecolor=blue,urlcolor=blue]{hyperref}

\begin{document}
\title{ Effect of quintessence on the Nature  of Kerr-newman blackhole shadow with clouds of strings}
\author{Gowtham Sidharth M}
\affiliation{School of Advanced Sciences,Vellore Institute of Technology, Chennai - 600048, India}

\author{Sanjit Das}
\affiliation{School of Advanced Sciences,Vellore Institute of Technology, Chennai - 600048, India}

\date{\today}

\begin{abstract}
In this paper we have took reissner nordstrom blackhole with cloud of strings and surrounds it with quintessence. we processed the metric through newman janis algorithm to get its rotating counterpart. the blackhole in study now is a roating charged blackhole with clouds of string surrounded by quintessence. we studied its nature of effective potential and unstable photon orbits. Finally we have plotted the blackhole shadow for various variable profiles.

\end{abstract}


\maketitle

\section{Introduction}
From their early theoretical framework  to the latest observerence of the blackhole \cite{key-1,key-2}.It has remained the most enigmatic object in the entire cosmos.General Relativity establishes that the space time at the near vicinity of the black hole is so warpped that the photon orbits around the blackhole in an unstable circular orbit. There are plenty of research has been already done on both rotating and non-rotating cases\cite{key-3,key-4,key-5,key-6,key-7,key-8,key-9,key-10,key-11,key-12,key-13}.where the shadow is a perfect circle in later case and distorted in the former due to the frame-dragging effect. A blackhole with presence of charge has also been studied. In addition to this there are black holes with an addisional dark energy quintessenisal field which was first proposed by kesilev. the quintessential field is present to explain the accleratated expansion of our universe \cite{key-14,key-15,key-16}.

In this paper, Instead of assuming point particles,we consider the the blackhole to be reside on on dimensional strings \cite{key-16,key-17,key-18,key-19,key-20,key-21,key-22,key-23,key-24,key-25,key-26}.A static spherically symmetric spacetime metric for this black hole, has been studied  in Ref. In our case we considered a Ressiner nordstrom blackhole in a stingy universe with added quintessential field. using Newman Janis algorithm we have arrived at a metric for rotating charge blackhole with string cloud in pressece of quintessential fieldFin\

This Section 2 we discuss about this Kerr-Newman-Kiselev blackhole with string cloud, Section 3 discusses about the photon orbit and  effective potential,we have plotted the blackhole shadow in section 4 and summarized the paper in section 5.

\section{Kerr - Newman - kiselev BH with string cloud}

The metric of Reissner-Nordstrom surrounded by quintessence and cloud of string\cite{key-28} is given as,
\begin{equation}
\begin{split}
ds^{2}&=-\left(1-b-\frac{2M}{r}+\frac{q^{2}}{r^{2}}-\frac{c}{r^{3\omega_q+1}}\right)dt^{2}\\&+\left(1-b-\frac{2M}{r}+\frac{q^{2}}{r^{2}}-\frac{c}{r^{3\omega_q+1}}\right)^{-1}dr^{2}+r^{2}d\Omega^{2}
\end{split}
\end{equation}
Following newman janis algorithm\cite{key-27}, we have our metric to be,
\begin{equation}
\begin{split}
ds^{2}&=-\left(1-\frac{2\rho r}{\Sigma}\right)dt^{2}+\frac{\Sigma}{\Delta( r)}dr^{2}+\Sigma d\theta^{2}\\&-\frac{4 a \rho  r sin^{2}\theta}{\Sigma}{dtd\phi}+sin^{2}\theta\left(r^{2}+a^{2}+\frac{2a^{2}\rho r sin^{2}\theta}{\Sigma}\right)d\phi^{2}
\end{split}
\end{equation}
Where,
\begin{equation}
2\rho r\ =r^{2}+a^{2}-\Delta( r)
\end{equation}

\begin{equation}
\Delta( r)=\ \ r^{2}\ +\ a^{2}\ -\ 2\ M\ r\ -\ q^{2}\ +\ c\ r^{1-3\omega}\ +b\ \ r^{2}
\end{equation}

\begin{equation}
\Sigma = r^{2} + a^{2}\cos^{2}{\theta}
\end{equation}
In the above equations,q being the charge of the blackhole, M and
c term corresponds to mass and quintessence parameter,while a and
b denotes the spin and string cloud parameter. The metric reduces
to kerr newman kiselev blackhole when b = 0 and can be further reducedto
kerr newman with b = c = 0 and to kerr with b=c=q=0 and finally to
schwarschild by a=b=c=q=0.

Solving $\Delta( r)$ = 0 will give us the event horizon
, the plot $\Delta(r)$ vs r below gives us the comparison
for different blackholes.

\begin{figure}[H]
\includegraphics[scale=1]{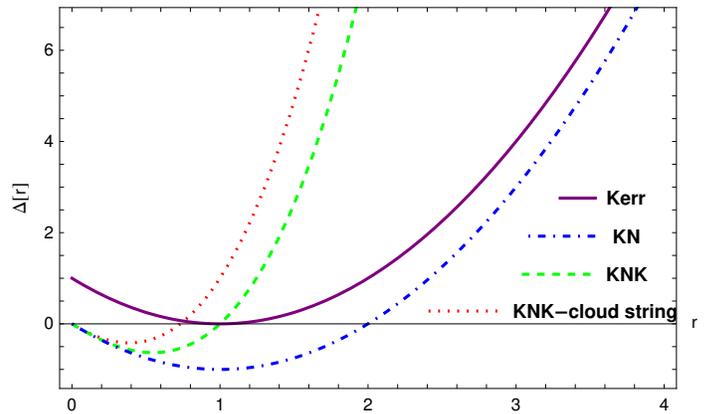}
\caption{Horizon plots for different blackholes.}
\end{figure}
\section{Photon Orbits}
\subsection{Unstable Circular Orbit}
There is a particular unstable orbit in which a photon revolves around
the blackhole,depending on the conditions of impact parameter the
would leave this orbit to either fell into the blackhole or escape
to infinity. The equation of such orbit can calculated using hamilton-Jacobi
Variable separation method\cite{key-29}. Hamilton - Jacobi equation in its general
form is given as,
\begin{equation}
\frac{\partial S}{\partial\lambda}= - \frac{1}{2}g^{{\mu \nu}}\frac{\partial S}{\partial x^{\mu}}\frac{\partial S}{\partial x^{\nu}}
\end{equation}
Where $\lambda$ is the affine parameter and S is the action thats defined
as,
\begin{equation}
S\ =\frac{1}{2}m^{2}\ \lambda\ -\ Et\ +\ L\varphi\ +\ S_{r}\ (r)\ +\ S_{\theta}\ (\theta)
\end{equation}
where m is the rest mass, E and L are energy and Angular momentum
which are the constants of motion.

Solving the Hamilton jacobi equation will yield,
\begin{equation}
\frac{\partial S_{r}}{\partial r}=\frac{\sqrt{R(r)}}{\Delta(r)}
\end{equation}

\begin{equation}
\frac{\partial S_{\theta}}{\partial\theta}=\sqrt{\Theta(\theta)}
\end{equation}
Where ,\\
 \strut \\
\begin{equation}
R(r)=(aL\ -\ (a^{2}\ +\ r^{2}\ )E)^{2}\ -\ (K\ +\ (aE\ -\ L)^{2}\ )\mathrm{\Delta}
\end{equation}
\begin{equation}
\Theta(\theta)=K\ -\left(\frac{L^{2}}{\text{si}n^{2}\theta}\ -\ a^{2}\text{\ E}\ ^{2}\right)Cos\theta
\end{equation}
where K is the the seperation constant. The trajectory of a photon
is calculated with two impact parameter,

$$\xi=\frac{L}{E}$$
$$\eta=\frac{K}{E^{2}}$$

Rewritting R(r) and $\Theta (\theta)$ in terms of impact parameters
\begin{equation}
R_{p}(r)=(a\ \xi\ -\ (a^{2}\ +\ r^{2}\ ))2-(\eta\ +\ (a-\xi)^{2}
\end{equation}
\begin{equation}
\Theta_{p}(\theta)\ =\ \eta\ - \left( \frac{\xi^{2}}{\text{si}n^{2}\theta}-\ a^{2}\ \right) cos^{2}\theta
\end{equation}
The equation of geodesic motion can be written the form
\begin{equation}
\Sigma\ t'=\frac{((r^{2}+\ a^{2}\ )E\ -\ aL)(r^{2}\ +\ a^{2}\ )}{\Delta(r)}a(aESin^{2}\ (\theta\ )\ -\ L)
\end{equation}
\begin{equation}
\Sigma\ r'\ =\sqrt{R(r)}
\end{equation}
\begin{equation}
\Sigma\ \theta'= \sqrt{\Theta (\theta)}
\end{equation}
\begin{equation}
\Sigma\ \varphi'=\frac{a((r^{2}\ +\ a^{2}\ )E\ -\ aL)}{\Delta(r)}-\frac{\text{aESi}n^{2}\ (\theta\ )\ -\ L}{\text{Si}n^{2}\ (\theta\ )}
\end{equation}
\section{Effective Potential}

The relationship between r' and effective potential is given as,
\begin{equation}
V_{\text{eff}}=\frac{E^{2}-1}{2}-\frac{1}{2}r'
\end{equation}
Effective potential is the maximum potential energy for a photon to
have zero radial velocity and zero radial acceleration and there by
determining the photons's stability.
\begin{figure}[H]
\label{v2}
\includegraphics[scale =1]{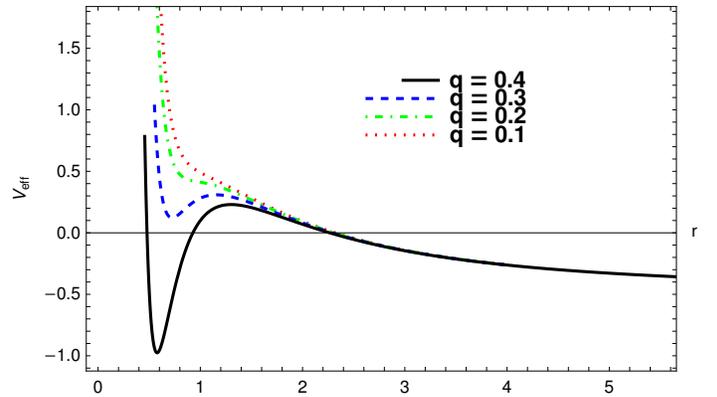}
\caption{$ V_{eff}$ vs r for varying q with M = 1; e = 1; k = 4; L = 2.1; a = 1; c = b = 0.1;.}
\end{figure}
\begin{figure}[H]
\label{3}
\includegraphics[scale = 1]{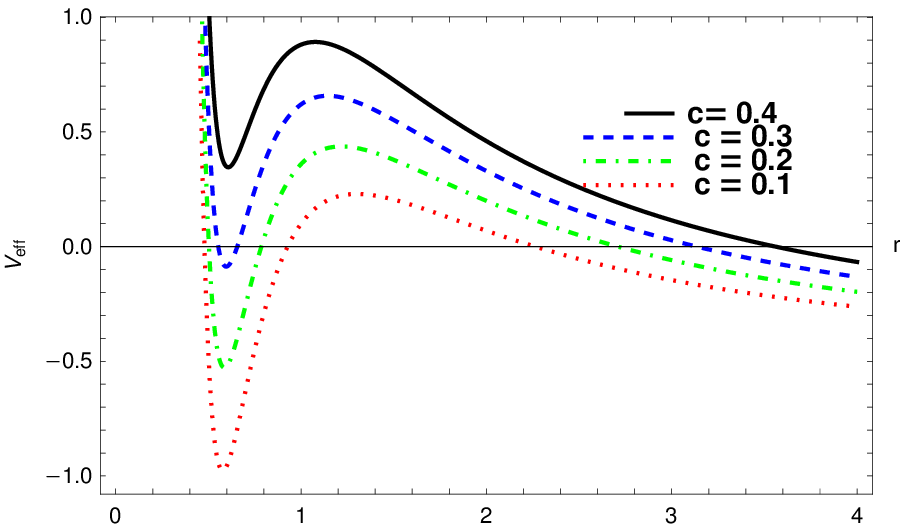}
\caption{$V_{eff}$ vs r for varying
c with M = 1; e = 1; k = 4; L = 2.1; a = 1; b = 0.1;q=0.4.}
\end{figure}

\begin{figure}
\label{v4}
\includegraphics[scale=1]{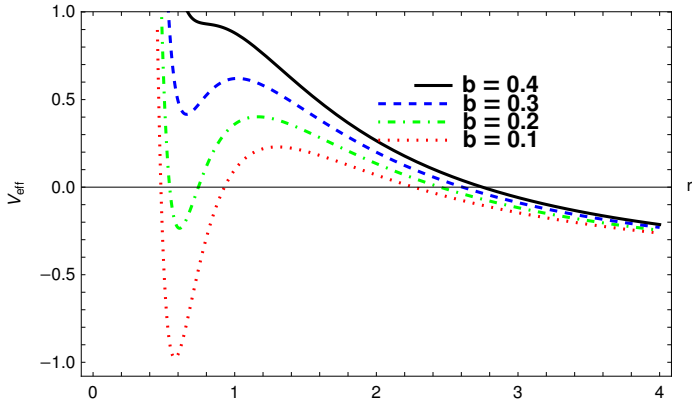}
\caption{ $V_{eff} $vs r for varying
b with M = 1; e = 1; k = 4; L = 2.1; a = 1; c = 0.1; q=0.4.}
\end{figure}
From figure 2 - 4 , we have plotted effective potential against r
for various profiles such as by varying charge 1, quintessence parameter
c and string parameter b. In fig 2, the maxima of the effective potential
decreases with increase in charge, we can also notice that the peak
is shifted toward the left as charge is being increased. In Fig 3,
the maxima increases with increase in quintessense value and peak
is shifted left. Similarly in fig 4, the maxima increases with string
parameter b and peak shifting towards to left can be seen here too.

\section{Shadows}

For a photon to have spherical orbit,it should have null radial velocity
and null radial acceleration,which means that,

$$R_{p}(r)=\ 0$$ 
$$\frac{\text{d\ }R_{p}}{\text{d\ r}}\ =\ 0$$

With this we can rewrite the impact parameter as,
\begin{equation}
\xi=\frac{-4r\mathrm{\Delta}(r)\ +\ a^{2}\ \mathrm{\Delta}'(r)\ +\ r^{2}\ \mathrm{\Delta}'(r)}{a\ \mathrm{\Delta}'(r)}
\end{equation}
\begin{equation}
\eta=\frac{r^{2}\ (16a^{2}\ \mathrm{\Delta}(r)\ -\ 16\mathrm{\Delta}^{2}(r)\ +\ 8r\mathrm{\Delta}(r)\ \mathrm{\Delta}'(r)\ -\ r^{2}\ \mathrm{\Delta}'^{2}(r)\ )}{a^{2}\mathrm{\Delta}'^{2}(r)}
\end{equation}
The Shape of the black hole depends on the celestial coordinates $\alpha$
and $\beta$ , which is given as
\begin{equation}
\alpha   = -{r_{0}}^{2}Sin \theta\frac{d\phi}{dr} 
\end{equation}
\begin{equation}
 \beta=\ -{{r_{o}}^{2}\frac{d\theta}{\text{dr}}}_{} 
\end{equation}
Where $\alpha$ and $\beta$ has limit $r_{o}$ goes to infinity,using
the geodesic equation the celestial coordinates can be written in
terms of impact parameter.
\begin{equation}
\alpha=\frac{-\xi}{sin(\theta)}
\end{equation}

\begin{equation}
\beta=\sqrt{\eta+a^2Cos^2(\theta)-\xi^2 cot^2(\theta)}
\end{equation}

The shadow can we obtained by plotting $\alpha$ vs $\beta$ in the equatorial plane.
\begin{figure*}
\includegraphics[scale=0.8]{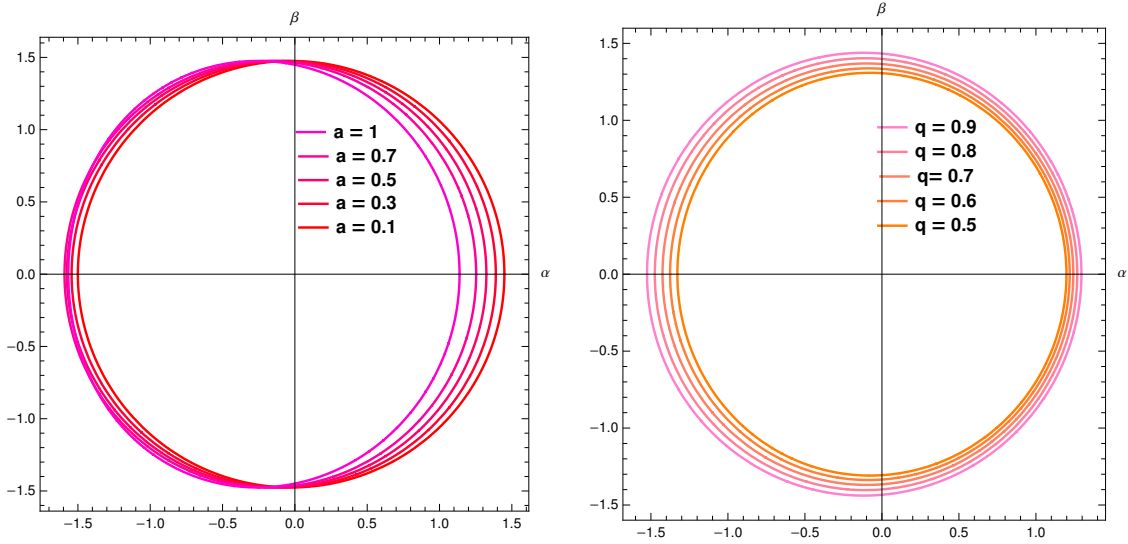}
\caption{Shadow plot fot KNK - cloud string black hole for
variation in spin parameter(left) snd charge(right)}
\end{figure*}

\begin{figure*}
\includegraphics[scale=1]{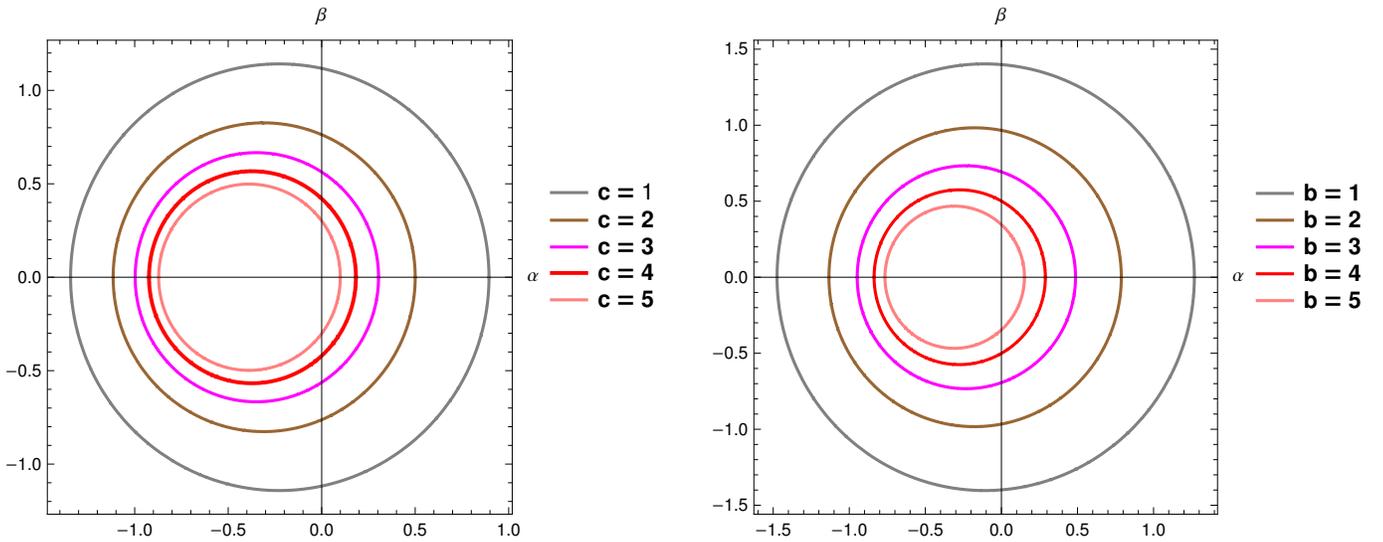}
\caption{ Shadow plot fot KNK - cloud string black hole for
variation in quintessense (left) snd string parameter(right).}
\end{figure*}
\section{conclusion}
In this paper we have took reissner nordstrom blackhole with cloud
of strings and surrounds it with quintessence. we processed the metric
through newman janis algorithm to get its rotating counterpart. the
blackhole in study now is a roating charged blackhole with clouds
of string surrounded by quintessence. we studied its nature of effective
potential and unstable photon orbits. we have observed that the effective
potential maxima is directly proportional to the increase in charge,quintessence
and string parameter and the peak seems to incline towards the left.

Finally, we have plotted the blackhole shadow for various variable
profiles.from figure 5, As expected the shadow shape starts to distort
as we go for higher spin values. it also seems to a distorted circles
for higher values of charge but upon closer observation,the circles
are in deed concentric proving that only size of the shadow decreases
with increase in charge. From figure 6, It is evident that the size
of the shadow decreases with increase in quintessence and string parameters.

\end{document}